# SOME NEW EXPERIMENTAL PHOTONIC FLAME EFFECT FEATURES

N.V.Tcherniega
P.N.Lebedev Physical Institute, RAS
Leninskii pr., 53, 119991, Moscow, Russia
tchera@sci.lebedev.ru

## Abstract

The results of the spectral, energetical and temporal characteristics of radiation in the presence of the photonic flame effect are presented. Artificial opal posed on Cu plate at the temperature of liquid nitrogen boiling point (77 K) being irradiated by nanosecond ruby laser pulse produces long- term luminiscence with a duration till ten seconds with a finely structured spectrum in the the antistocks part of the spectrum. Analogous visible luminescence manifesting time delay appeared in other samples of the artificial opals posed on the same plate. In the case of the opal infiltrated with different nonlinear liquids the threshold of the luminiscence is reduced and the spatial disribution of the bright emmiting area on the opal surface is being changed. In the case of the putting the frozen nonlinear liquids on the Cu plate long-term blue bright luminescence took place in the frozen species of the liquids. Temporal characteristics of this luminiscence are nearly the same as in opal matrixes.

# 1. Introduction

In [1-3] new effect called photonic flame effect was found and some its properties were studied. This effect is determined by properties of photonic crystals.

Photonic crystals have attracted great attention since the first papers concerning such structures [4-6]. One of the most important photonic crystals are artificial opals – self – assembled structures composed of $SiO_2$ spheres organizing face-centered cubic lattice. The size of such spheres varying between 200 nm and 400 nm and defines the parameters of the face-centered cubic lattice and the photonic bandgap. The possibility of opal infiltration with different medium gives rise to effective processing the properties of the light propogating through the crystal. The voids in the opal structures can be filled with semiconductors, superconductors, ferromagnetic materials, fluorescent medium [7] and this fact gives large possibility to practical applications of such structures for optoelectronics. The study of the linear optical properties of the photonic band gap have been the task of many theoretical and experimental works and still remain the task to be investigated [7,8]. The theoretical description of the electromagnetic field inside the photonic crystal structures (obtained by transfer matrix method [9] or coupled mode theory [10]) gives the clear picture of the transmitted and reflected spectrum, electromagnetic field distribution inside the crystal and their dependence on the parameters of the photonic crystal structure (values of period, number of periods, refractive index contrast). Large values of the electromagnetic field localization in some regions lead to the expectation of the strong enhancement of nonlinear wave-matter interaction in comparison with bulk crystals. Second harmonic generation in different types of photonic crystals was investigated in [11,12]. Properly chosen photonic crystal exhibits negative refraction at some conditions [13]. Some features of the stimulated Raman scattering in one-dimensional photonic structure were considered in [14]. Fully quantum mechanical treatment of the generation of entangled photon in nonlinear photonic crystals at the process of down-conversion was realized in [15]. Photonic band gap properties which are demonstrated by photonic crystalls are being actively used for investigation of photon-exciton interaction [16]. Acoustic modes excited in $SiO_2$ balls which compose opal photonic crystal show the effect of phonon modes quantization [17] and are the reason of stimulated globular scattering [3]. Specific features of the acoustic wave propagation in the photonic structures lead to the possibility of the diverging ultrasonic beam focusing into a narrow focal spot with a large focal depth [18]. Optical parametric oscillations via four-wave mixing in isotropic photonic crystals showes the possibility of the effective frequency processing [19].

The aim of this work is to give a short review of results [1-3] and to study collective behavior of several photonic crystals. The crystals are posed on Cu plate at the temperature of liquid nitrogen. One of the photonic crystals is illuminated by laser pulse and the laser light is focused on this only crystal. The phenomenon which we observe is the appearance of luminescence of other photonic crystals. The duration of the luminescence of other crystals which are spatially separated with the crystal illuminated by laser pulse is of the order of seconds. The appearance of the luminescence takes place with some time delay respectively to the laser pulse. The form of these light spots on the other crystals and their slow motions along the crystal reminds a small flame spot. This inspired us to give the "photonic flame" name to the observed effect. In the case of covering the surface of the Cu plate with liquid (acetone, ethanol, water) after the PFE excitation in the opal situated on this plate blue luminescence is being seen in the frozen liquid. The temporal characteristics of this luminescence are the sme as for single opal crystal. The paper is organized as follows. In Sec.2 the experimental setup, laser, the photonic crystals (artificial opals) used in the experiment are described. In Sec.3 the "photonic flame effect" observed in the experiment is discussed. In Sec.4 perspectives and possible explanations are presented.

## 2. Photonic crystals and laser used in experiment.

One of the most promising three-dimensional photonic crystals is artificial opal. Opal is a crystal with face-centered cubic lattice consisting of the monodisperse close packed $SiO_2$ spheres with diameter about several hundred nanometers. Because the refractive index contrast (ratio $n_{SiO2}/n_{air}$) is about 1,45 the complete photonic band gap does not exist but the photonic pseudogap takes place. Empty cavities among these globules have octahedral and tetrahedral form. It is possible to investigate both initial opals (opal matrices) and nanocomposites, in which cavities are filled with organic or inorganic materials, for instance, semiconductors, superconductors, ferromagnetic substances, dielectrics, displaying different types of nonlinearities and so on. Filling voids of the photonic crystal with materials with different refraction index one can effectively process the parameters of the photonic pseudogap.
Ruby laser giant pulse ($\lambda$=694.3 nm, $\tau$=20 ns, $E_{max}$ =0.3 J, spectral width of the initial light - 0.015 cm$^{-1}$.) has been used as a source of excitation. Exciting light has been focused on the material by lenses with different focal lengths (50, 90, and 150 mm). The samples of opal crystals used had the size 3x5x5mm and were cut parallel to the plane (111) (see Fig.2) .The angle of the incidence of the laser beam on the plane (111) varied from 0 to 60$^0$. Sample distance from focusing system and exciting light energy were different in different runs of the experiment. This gave possibility to make measurements for different power density at the entrance of the sample and for different field distribution inside the sample. Opal crystals consisting of the close-packed amorphose spheres with diameter 200 nm, 230 nm, 250 nm and nanocomposites (opal crystals with voids filled with acetone or ethanol) were investigated.

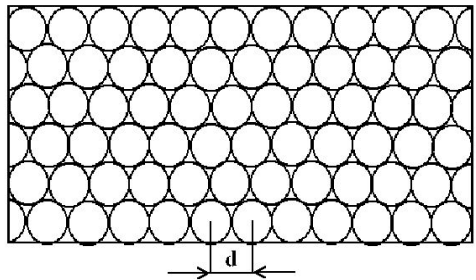

Fig.1. Common appearance of a globular photonic crystal, built of spherical particles (globules)

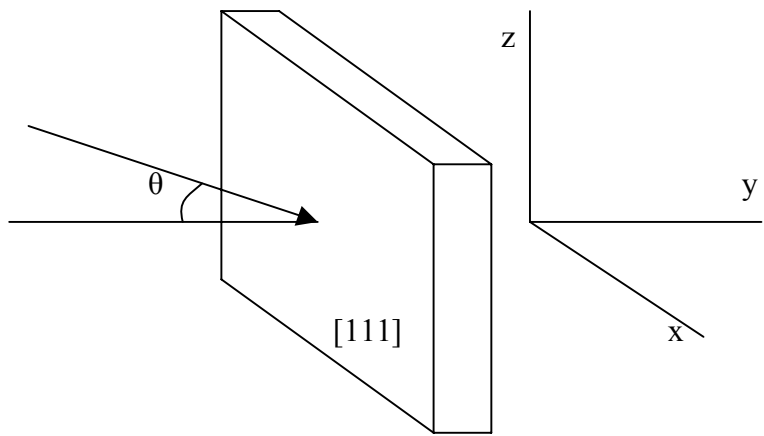

Fig.2. The scheme of illuminating the sample. Plane XY correspondes to the CU plate surface.

## 3.Characteristics of "photonic flame"

Opal crystals were placed on the Cu plate which was put into the cell with liquid nitrogen (see Fig.3). The number of crystals varied from 1 to 5. The distance (d) betwen the crystals was of the order of several centimeters (maximum value of d was 5 centimeters and was determined by the Cu plate size). One of the crystals was illuminated by the focused laser pulse. In the case of the reaching of the threshold visible (blue) luminescence appeared. The luminescence duration was from 1 to 12 seconds and it looked like inhomogeneous spot changing its spatial distribution and position on the surface of the crystal during this time.

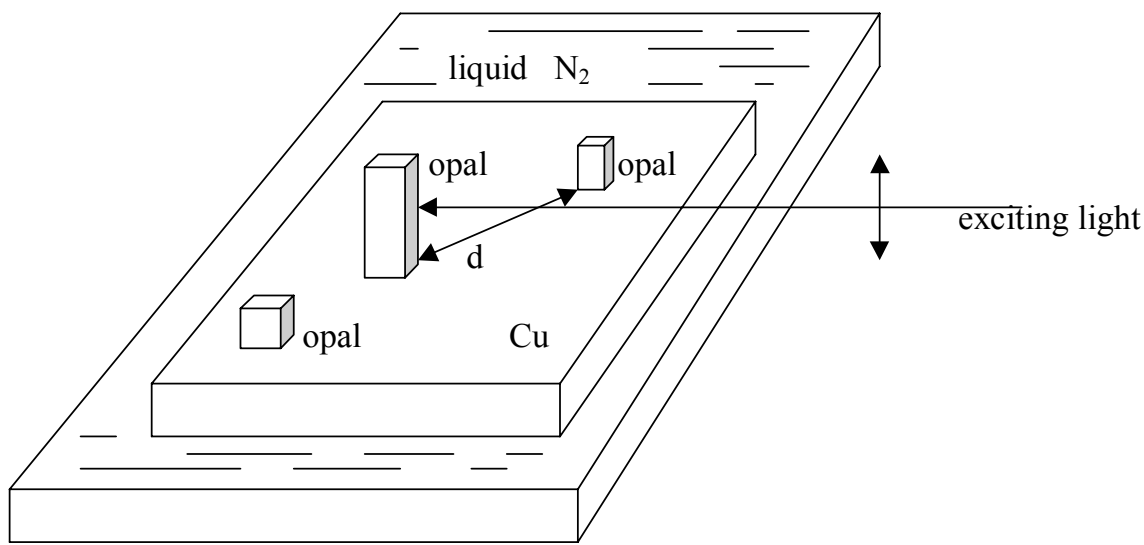

Fig.3. Experimental setup.

Parameters of the luminescence (duration, threshold) were determined by the geometric characteristic of the illumination and the refractive index contrast of the sample. For optimal geometry of the excitation the power density threshold for opal crystal was 0.12 Gw/cm$^2$, for opal crystals filled with ethanol – 0.05 Gw/cm$^2$, for opal crystal filled with acetone - 0,03 Gw/cm$^2$. Typical luminescence temporal distribution measured for the part of the crystal displaying the most intensive brightness is shown on Fig.4. The same behavior is typical for all cases of the luminescence at these conditions of excitation, but the value of the luminescence duration fluctuated from shoot to shoot.

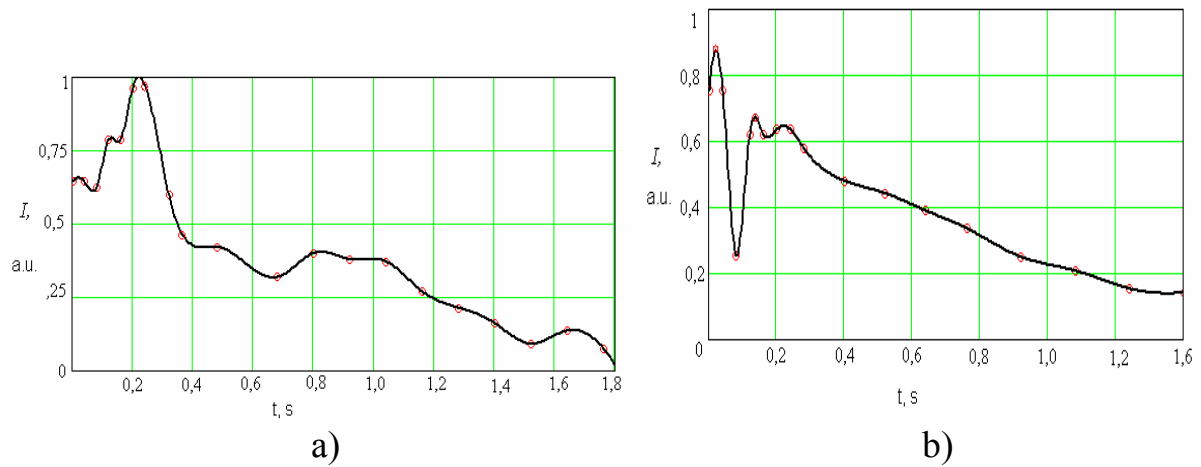

a) b)

Fig.4. Temporal distribution of the visible luminiscence.

The duration of the luminiscence fluctuated from 1 till 12 seconds and demonstrated oscillating structure. In some cases the temporal distribution had maximum at the beginning of the luminiscence in some cases – minimum. Fig.4 a) and b) show the luminiscence of the pure opal matrix of the nearly the same duration at the same geometrical and energetical conditions of excitation near the threshold of excitation (0.12 Gw/cm$^2$ ). The beginning of the mesurements corresponds to 0.3 s delay after the laser shoot (laser pulse duration is 20 ns).

Secondary emission spectrum observed in photonic flame effect has been investigated with the help of setup shown at the Fig.5.

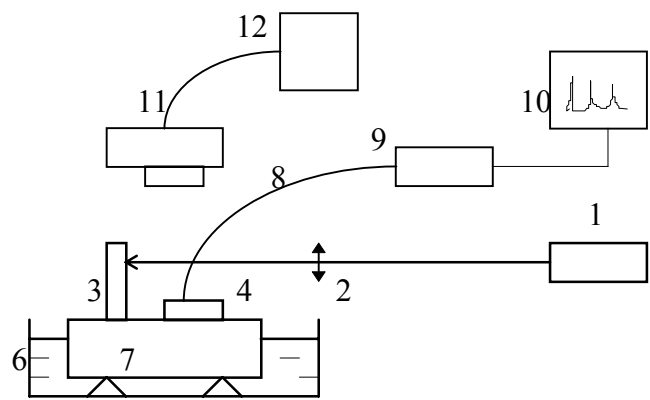

Fig. 5.The experimental setup for PPE spectrum study. 1- ruby laser; 2- lens; 3, 4, 5 – photonic crystals; 7 – cell with liquid nitrogen 6 – Cu plate; 8 – fiber wave guide; 9 – minipolychromator; 10 – computer; 11 – camera; 12 – computer.

Spectra of the light emitted by photonic crystal for different pumping light power density are shown at the Fig. 6 (a and b)

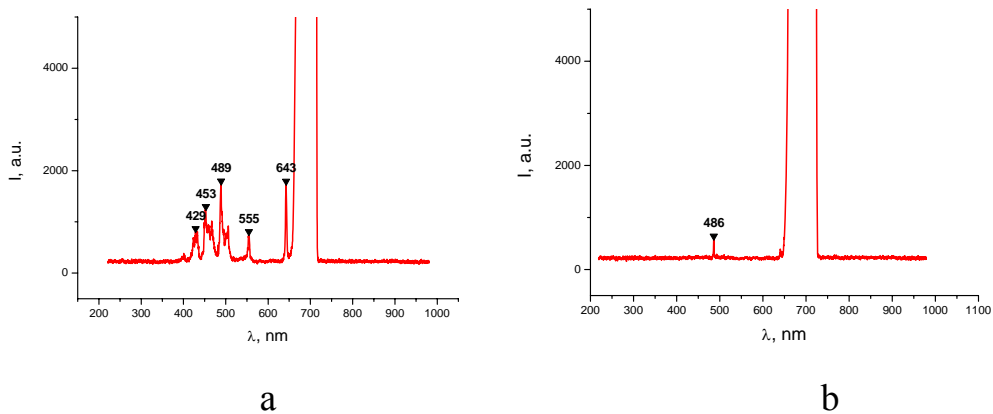

a  b

Fig. 6. Secondary emission spectrum of a photonic crystal for different laser light power density:
a - I = 0.12 GW/cm$^2$,   b - I = 0.14  GW/cm$^2$.

Spectrum consisted of the sharp lines with wavelengths: 429.0, 453.0, 489.0, 555.0, 643.0 nm, which corresponds to the antistokes spectral range for exciting line 694.3 nm. Lines intensity in the spectrum strongly depended on the laser pumping intensity, which was evidence of stimulated type of the radiation emission.

In the case of several crystals placed on the Cu plate only one of them was irradiated by the laser pulse. Luminescence took place in this crystal in the case of the threshold reaching. Bright shining of the other crystals began with some time delay after laser shoot. The value of this delay (and the intensity of the luminiscence) was determined by the spatial position of the crystals on the plate. The steal screen beeing put between the crystals (in order to avoid irradiating of the crystals by the light scattered by the crystal excited by the laser) did not stop the appearing of the luminescence if the distance between the Cu plate and the screen was more than 0.5 mm. The duration of the luminiscence was of the order of several seconds and temporal behavior was like shown on Fig.4. The typical features of such distribution were existence of maximum and large plato with near constant value of the intensity.

In order to show the role of the material of the plate used we repeated these measurements with plates of the same size but made from steel and quarz on which opal crystals were placed like in the previous experiments. Luminescence of the same kind in the irradiated crystal took place but the luminescence of the other samples situated on these plates was not observed.

The effect was also determined by the angle of incidence (Fig.2). For the samples used the value of the angle was chosen experimentally for achieving of the maximal value of the luminescence (it worth to mention that this value differed from 0 and was about 40$^0$). Easier the effect was excited in the unprocessed samples. In Fig.5 one can see the luminescence of the crystals situated at the distance of about 1 centimeter from the crystal which was irradiated.

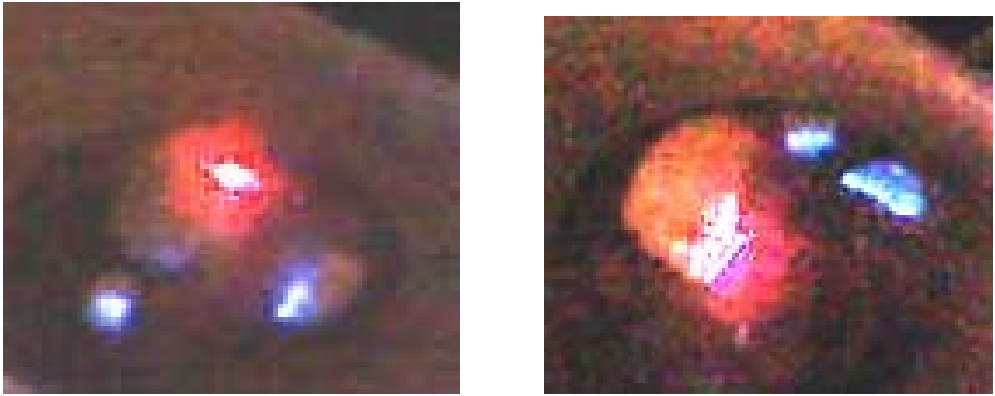

Fig.7 Visible luminiscence of the opal crystals in the case of the irradiating one of them (the irradiated crystal can be seen by bright red light; on the left picture it was the crystal in the center, on the right picture it was the crystal on the left). Left picture corresponds to the case where crystals are infiltrated by acetone. Right picture corresponds to the case of the opal crystals without infiltration.

In the case of the large laser energy (several times more the threshold) or if the crystal was irradiated by several laser pulses the opal can be destroyed and the parts of the crystal produce the luminiscence with the spectral and temporal properties described above (Fig.8).

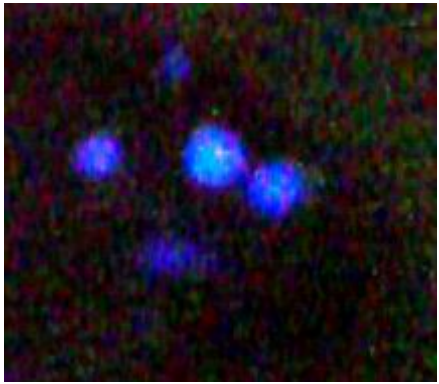

Fig.8 Opal crystal is destroyed and 3 large pieces and several little pieces are going on to produce the luminescence.

In order to clarify the role of the Cu plate surface on the energy transport between the crystals the next experiment was realized: the pure opal matrix posed on the Cu plate was irradiated by the ruby laser pulse and demonstrated strong luminescence lasting few seconds with the properties described above (Fig.9)

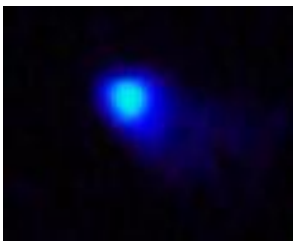

Fig.9 Luminiscence of the single opal matrix

Next step was covering the surface of the Cu plate with the liquid (experiments were made with water, aceton or ethanol). The thickness of the frozen liquid on the plate surface was about 1 mm. The transverse size of the frozen liquid was about 1 cm. After illuminating of the crystal by the ruby laser pulse the luminiscence of the crystal appears the bright blue luminiscence of the frozen species of the liquid used appears. The temporal characteristics of the luminiscence in crystal and in the frozen liquid are approximatly the same (the luminiscence duration is about several seconds). The luminiscence of the frozen liquid goes on in spite on the putting the screen between the crystal and the liquid. It shows that the luminiscence of the liquid is not a reflection of the light which is emmited by the crystal. Fig.10 shows the luminiscence of the crystal and the frozen liquid (in this case it was water). The pictures were made with the interval of 1 second between each other. Analogous behaviour is demonstrated by aceton and ethanol. The luminiscence of the area covered with frozen liquid takes place even if this area is at the distance of several cantimeters from the irradiated crystal. The explanation of the blue luminiscence of the frozen liquid can be done in several ways and for clarifying the reasons of this luminescence appearance it is necessary to produce additional experiments. The intensity of the laser in the experiments is about 0.12 $Gw/cm^2$, and the large enhancement of this field due to Mie – resonance [20] simultaniously with the interference effect caused by the structure of the opal matrix can lead to the extremely large field enhancement which can play an important role in this effect.

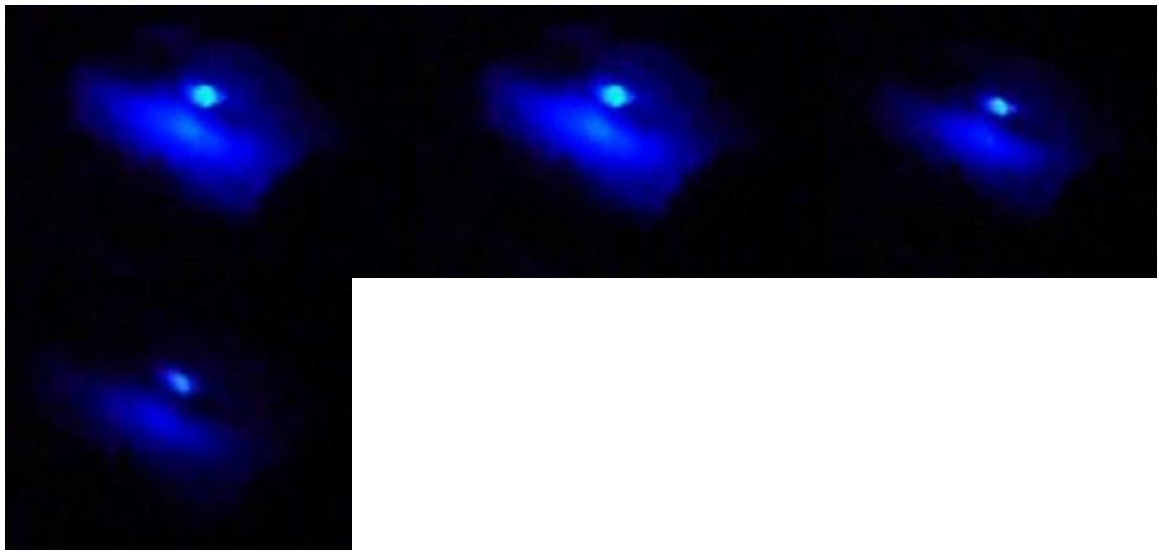

Fig.10 Luminiscence of the opal matrix (bright round spot) and frozen liquid (large blue spot) on the surface of the Cu plate.

## 4. Conclusions.

In this paper we reported about some new features of the photonic flame effect. The main features of PFE are:
- At the excitation of the artificial opal crystal which is placed on the Cu plate at the temperature of the liquid nitrogen by the ruby laser pulse of the nanosecond range long-

- continued optical luminescence takes place in the case if the threshold of the process is reached;
- In the case of several opal crystals being put on the Cu plate while one of them is being irradiated bright visible luminescence occurs in all samples;
- Temporal behavior and thresholds of the luminescence have been determined. Photonic crystals infiltrated with different nonlinear liquids and without infiltration have been investigated. Investigated transport of the excitation between the samples spatially separated by the length of several centimeters gives the possibility of the practical applications of PFE;
- The blue luminiscence of the frozen liquid on the surface of the Cu plate takes place at the precense of the photonic flame effect;

The photonic flame effect can have different explanation. Probably an essential role is played by plasma properties. The slow transport of the excitations from the irradiated crystal to other photonic crystals can be associated with sound waves created due to laser pulse interaction with the sample. Exciton mechanism and surface waveguides on the surface of the Cu plate also can play important role. It was checked that the change of the properties of the plate surface was leading to change of the photonic flame effect. Removing the oxid layer from the plate changed the threshold PFE. The luminescence of the frozen liquids on the surface of the Cu plate showes the important role of the electromagnetic field enhancement due to Mie resonance and Bragg diffraction on the photonic crystal lattice. The electromagnetic field enhancement can lead to producing laser plasma, electron acceleration and x-ray production.